\documentstyle[prl,aps,psfig,twocolumn]{revtex}

\begin{document}
\draft 
\input amssym.def
\input amssym.tex
%
%
\def\ds{\displaystyle}
\def\e{\varepsilon}
\def\FigSize{8.40cm}\def\FigJump{0.20cm}\def\LOC{ht} 	
\def\root{}
\def\p{\partial}
\def\pp#1#2{{{\p x^{n+1}_{#1}}\over{\p x^{n}_{#2}}}}
%
\def\oneFIG#1#2#3{\begin{figure}[\LOC]
\centerline{\psfig{figure=\root#1,width=#2,silent=}}
\vskip \FigJump \caption[]{#3\label{#1}} \end{figure} }
%
\def\twoFIG#1#2#3#4{\begin{figure}[\LOC]
\centerline{\psfig{figure=\root#1,width=#3,silent=}}
\vskip 0.2cm
\centerline{\psfig{figure=\root#2,width=#3,silent=}}
\vskip \FigJump \caption[]{#4\label{#1}} \end{figure} }
%
\def\CMLLOGLSCAP
{a) Lyapunov spectrum for a CML of $N=20$ fully chaotic logistic maps
$f(x) = 4x(1-x)$ with coupling strength $\e=0.4$. b) The Lyapunov dimension
$D_L$ and the KS entropy $h$ may be estimated from the sum of the Lyapunov
exponents. In this case the largest Lyapunov exponent is
$\lambda_1\simeq 0.36$, $D_L\simeq 15.4$ and $h\simeq 1.8$.
\par}
\def\CMLLOGDSONECAP
{Estimates of the LS for a logistic coupled
map lattice of size $N=20$ and coupling $\e=0.4$ computed using a time
delay reconstruction (with $d\in[32,42]$) from a univariate time series
(thin lines). A second order local fit from a sample of $10^4$ points was
used and
the number of neighbours was set to the number of parameters in
the fitting process plus 20. For comparison the spectrum computed from the
original dynamics is depicted with a thick line. An orbit length of $10^4$
iterations was used in both cases and the computed LS was rescaled using
(\ref{rescaling}).
\par}
\def\CMLLOGDSDTCAP
{LS estimated from spatio-temporal delay reconstructions for a
logistic coupled map lattice. The LS computed from the original dynamics is
depicted with a thick line while the different spatio-temporal delay
reconstructions are depicted with dashed lines.
All the fitting parameters are the same as in figure \ref{cmllogDS1F.ps}
(second order fitting, $10^4$ points, $10^4$ iterates).
\par}
\def\CMLLOGDTONECAP
{LS estimated from pure spatial delay reconstructions for a
logistic coupled map lattice.
All the fitting parameters are the same as in figure \ref{cmllogDS1F.ps}
(second order fitting, $10^4$ points, $10^4$ iterates).
\par}
\def\CMLLOGDTONECONECAP
{LS estimated from pure spatial delay reconstructions for a
logistic coupled map lattice using truncation of the outer layer of the
Jacobian. All the fitting parameters are the same as in figure
\ref{cmllogDS1F.ps}
(second order fitting, $10^4$ points, $10^4$ iterates).
\par}
\def\SKEWCAP
{Skewed logistic map (thick line) for for $b=0.2$  obtained by applying
(\ref{skew-transform}) to the fully chaotic logistic map (dashed line).
\par}
\def\SKEWBTWOCAP
{Estimation of the LS for a coupled map lattice of skewed logistic maps.
The size of the lattice is $N=20$, the coupling parameter is $\e=0.4$ and
the skew parameter is $b=0.2$. The original LS
is indicated by the solid line and the estimates using a range of spatial
embedding dimensions $d_s$ are depicted by various symbols.
For large $d_s$ the estimates deteriorate because of the
``curse of dimensionality''.
All the fitting parameters are the same as in figure \ref{cmllogDS1F.ps}
(second order fitting, $10^4$ points, $10^4$ iterates).
\par}
\def\SKEWSTWENTYCAP
{Effect of the skew of the local map when estimating the LS.
Curves correspond to the LS computed with the known dynamics and
the symbols depict various  pure spatial
delay reconstructions.
All the fitting parameters are the same as in figure \ref{cmllogDS1F.ps}
(second order fitting, $10^4$ points, $10^4$ iterates).
\par}
\def\SKEWNTCAP
{Effect of truncating the outer row of the Jacobian for a
tri-diagonal estimate of the LS for a lattice of coupled skewed
logistic maps as in figures \ref{skewDT1b2F.ps} and \ref{skewDT1s20F.ps}.
The skew is $b=0.2$ and the width of the window used to extract
the multivariate time series is $d_s$ sites. The open symbols correspond
to using the whole fitted Jacobian while the solid symbols correspond
to truncating the outer layer of the Jacobian.
All the fitting parameters are the same as in figure \ref{cmllogDS1F.ps}
(second order fitting, $10^4$ points, $10^4$ iterates).
\par}
\def\SKEWQCAP
{Tri-diagonal estimate of the LS for a lattice of coupled skewed
logistic maps as in figures \ref{skewDT1b2F.ps} and \ref{skewDT1s20F.ps}.
The skew is $b=0.2$ and the width of the window used to extract
the multivariate time series is 20 sites.
All the fitting parameters are the same as in figure \ref{cmllogDS1F.ps}
(second order fitting, $10^4$ points).
\par}
\def\EXPCOUPCAP
{Estimation of the LS using a $(2q+1)$-diagonal reconstruction of the
Jacobian for a lattice of coupled skewed logistic maps with decaying
exponential coupling (\ref{exp-coup}). The skew is $b=0.2$
and the decay rate is $\beta=0.35$. The coupling parameters
for $\beta=0.35$ correspond to $\e_0\simeq 0.481$, $\e_{\pm 1}\simeq 0.168$,
$\e_{\pm 2}\simeq 0.059$, $\e_{\pm 3}\simeq 0.021$,
$\e_{\pm 4}\simeq 0.007$, \dots, in the general CML formulation
(\ref{gral-CML}).
All the fitting parameters are the same as in figure \ref{cmllogDS1F.ps}
(second order fitting, $10^4$ points, $10^4$ points).
\par}
\def\MESHPDECAP
{Cone-horizon for a disturbance in a one-di\-men\-sio\-nal, space-time
continuous, extended dynamical system. A disturbance applied at the summit of
the cone may alter the downstream dynamics inside the cone-horizon (shaded
area). Applying a given space-time discretisation (dashed mesh) induces a
particular neighbourhood structure: a node at time $t+\tau$ depends on a
fixed number of nodes at time $t$ (3 in this case). The slope of the
cone-horizon fixes the number of sub-diagonals required in the
quasi-diagonal estimation of the Jacobians.
\par}

\wideabs{
\title{\bf A quasi-diagonal approach to the estimation of Lyapunov spectra \\
           for spatio-temporal systems from multivariate time series}
\author{
R.~Carretero-Gonz\'alez\cite{rcg:email},
S.~{\O}rstavik, and J.~Stark}
\address{Centre for Nonlinear Dynamics and its
         Applications\cite{CNDA:web},\\
         University College London, London WC1E 6BT, U.K.}

\date{PREPRINT VERSION, \today,
     check http://www.math.sfu.ca/$\sim$rcarrete/abstracts.html for updates}
\maketitle

\begin{abstract}
We describe methods of estimating the entire Lyapunov spectrum of a
spatially extended system from multivariate time-series observations.
Provided that the coupling in the system is short range, the Jacobian has a
banded structure and can be estimated using spatially localised
reconstructions in low embedding dimensions. This  circumvents the ``curse
of dimensionality'' that prevents the accurate reconstruction of
high-dimensional dynamics from observed time series. The technique is
illustrated using coupled map lattices
as prototype models for spatio-temporal chaos and is found to work even
when the coupling is not strictly local but only exponentially decaying.
\end{abstract}

}

\tableofcontents

\section{Introduction}

One of the most important tools for investigating chaotic dynamical
systems is the spectrum of Lyapunov exponents. These exponents measure
the asymptotic exponential divergence or convergence of two
infinitesimally close orbits. In this paper we are interested in
estimating {\em all} of the Lyapunov exponents of a spatio-temporal
system from an observed multivariate time-series without prior knowledge
of the dynamics governing the system. Hitherto, most efforts in this
area have concentrated on only estimating the largest (or few largest)
exponent(s). However, at least some of the negative exponents are needed
if we wish to estimate the dimension of the attractor via the
Kaplan-Yorke conjecture. Estimating all of the Lyapunov exponents from a
time series for spatially extended systems is a daunting task. The
fundamental problem is the high dimensionality of the system which
prevents an accurate reconstruction of the dynamics from observed data.
In particular, whatever method of function approximation we use to
reconstruct the dynamics, we need to have a reasonable
spread of data in the region of interest. Thus, for instance, a local
linear or quadratic approach estimates the value of a function at a
point $y \in {\Bbb R}^d$ by performing a linear least squares
regression in a neighbourhood of $y$. Such a neighbourhood must contain
a sufficient number of data points to yield a meaningful estimate and
yet not be so large that the function is no longer linear (or quadratic
respectively). As $d$ grows, more and more data is necessary to ensure
that sufficient numbers of neighbours can be found and for $d$ much larger
than 6 or so the amount of data required becomes completely impracticable.
This problem is often informally referred to as the ``curse of
dimensionality''.

It is therefore perhaps surprising that, at least in some cases, the whole
Lyapunov spectrum can be estimated successfully from observed 
data, see Ref.~\cite{bunner-hegger:99} and \cite{Sakse-rcg:99}. 
Both of these papers focus on a
lattice of locally coupled fully chaotic logistic maps, though
Ref.~\cite{Sakse-rcg:99} (here referred as ORS99)
also considers the effects of skewing the map (as we
also do below in \ref{LS-BASIS}), and of replacing it by a much more
non-linear function. The results for the latter two cases are substantially
less satisfactory than for the standard logistic map. It turns out that the
key property that makes it possible to obtain reasonable estimates of the
spectrum for the logistic map lattice using the methods in
ORS99 is that the dynamics at one spatial location is easily
approximated within the space of functions used to fit the dynamics. Hence
what at first sight appears to be a {\em local} quadratic fit is in fact a
{\em global} fit. This allows the ``curse of dimensionality'' to be
circumvented, and good estimates of the dynamics to be obtained even in
high dimensions.
As the local dynamics moves away from the space of functions used to fit
the dynamics the estimates of the Lyapunov spectrum rapidly deteriorate.
One approach to overcoming this might be to attempt to estimate the local
dynamics and then use a suitable basis to fit the dynamics\cite{rcg:loc-dyn}.

The alternative, which is the method used in Ref.~\cite{bunner-hegger:99} 
(here referred as BH99) and
which we pursue here, is based on the observation that as long as the
spatial coupling in the system is reasonably short range, optimal
predictions are often obtained in very low embedding dimensions\cite{Sakse:98},
even when the dimensionality of the attractor is high.
This somewhat counter-intuitive result may be explained by the rapid
spatial decay of the dependence of the dynamics at one site on its
neighbours\cite{rcg:trunca}. This suggests that it ought to be possible to
obtain reasonable estimates of the Jacobian by performing appropriate fits
in low embedding dimensions, hence avoiding the ``curse of dimensionality''.

The aim of this manuscript is to describe a method based on this intuitive
idea and evaluate its performance in various circumstances. The method is
based on only reconstructing the non-zero entries of the Jacobian in a
row-by-row fashion. More precisely if the coupling in a spatio-temporal
system is reasonably local then the only non-zero entries in the Jacobian
are near the diagonal. Rather than estimating the whole Jacobian, one
should only estimate the non-zero entries (which can be done in a low
embedding dimension) and set the remaining entries to 0. We call such an
approach a {\em quasi-diagonal} one. Essentially the same technique is used
by B\"unner and Hegger in BH99 who successfully apply it to
the logistic coupled map lattice, using local linear fits. However, as
ORS99 shows, the standard logistic lattice is a relatively
easy system for which to estimate the spectrum, even using local linear
fits (rather than quadratic ones, as in ORS99 and here). It
is thus impossible to judge from BH99 how well a
quasi-diagonal approach works when the local dynamics presents a bigger
challenge, or indeed when the coupling is anything but nearest neighbour.
Finally, ORS99 demonstrates that dramatically improved
estimates of the spectrum can be obtained by truncating the outer layer(s)
of the estimated Jacobian, thereby eliminating boundary effects. Such
truncation is not considered in BH99.

The present work combines the best features of BH99 
and ORS99 together with additional generalisations and extensive
supporting numerical evidence. In particular, we evaluate the
quasi-diagonal approach when applied to a more difficult local map, using a
more flexible (quadratic) fitting basis. We investigate the effect of
truncating the outer layers of the estimated Jacobian and assess the
effectiveness of our approach when the coupling is exponentially decaying,
rather than just nearest neighbour. The encouraging numerical results we
obtain motivate us to present a possible extension of the
quasi-diagonal approach to continuous space-time extended dynamical systems
({\em i.e.}~partial differential equations).

For the convenience of the reader, we present our approach in a largely
self-contained manner, leading in a natural progression from a scalar
method, which gives very poor results, to a successful quasi-diagonal
estimation of the Jacobian from multivariate time series. In the
process we highlight the relationship between the ``globality'' of the
fitting basis and the ``curse of dimensionality''. We also emphasise the
importance of properly addressing the effects of the boundaries of the
subsystem where we collect the multivariate time series.

The paper is organised as follows.
%
%
The next section gives a general introduction to
spatio-temporal systems and describes the prototype model (a coupled
map lattice) that is used in our numerical investigations.
%
%
In section \ref{LS-KNOWNJ} we provide a short overview of
Lyapunov spectra for extended dynamical systems
and their relationship to fractal dimensions and the Kolmogorov-Sinai
entropy. We give some examples and explain how to estimate the Lyapunov
spectrum from subsystem information using a suitable rescaling.
%
%
In the following section (\ref{LS-UNIVARIATE}) we attempt to estimate the
Lyapunov spectrum using a scalar time series. We find that the lack of
spatial information hinders any attempt to obtain a meaningful
reconstruction of the dynamics, and hence to estimate the spectrum.
%
%
In section \ref{LS-MULTIVARIATE} we present
a systematic numerical study of estimates of the Lyapunov spectrum
using different spatio-temporal reconstructions. We discover the importance of
including spatial information in order to obtain reasonable
reconstructions of the dynamics. We also point out that boundary effects on
the measured subsystem have to be properly addressed
by truncating the outer layer(s) of the estimated Jacobian.
%
%
In section \ref{LS-BASIS} we turn our attention
to the effect of passing from a local fit to a global fit when
increasing the embedding dimension of the spatio-temporal reconstruction.
We conclude that the ``curse of dimensionality'' precludes any
hope of a usable reconstruction when our fitting basis does not
give a good global approximation to the dynamics.
%
%
We then go on in section \ref{LS-QUASIJ} to exploit the sparse structure of
the Jacobian when the coupling is short range to develop a quasi-diagonal
estimation technique for the Jacobian. Additionally, we use the spatial
homogeneity of the system to dramatically increase the amount of effective
data points available when performing a local fit. This circumvents both
the  ``curse of dimensionality'' and the error induced by not using an
appropriate global basis.
%
%
Section \ref{LS-GRALCOUPLING}
is devoted to the generalisation of our approach to systems
that have non-local, but exponentially decaying coupling.
%
%
Finally, in the last section, we propose a natural extension of our method
to the estimation of the Lyapunov spectra of partial differential equations.

\section{Spatially extended systems}

The occurrence of chaos in spatio-temporal systems has recently attracted
the attention of a large part of the dynamical systems community.
There exists nowadays a broad understanding of low-dimensional
chaotic systems. However, the same cannot be said of high-dimensional
systems and in particular of spatially extended systems.
The addition of a spatial extent to the dynamics produces a complex
interplay between the local dynamics (the original dynamics before
including spatial interactions) and the spatial interactions.
Sometimes, this interplay triggers the so called phenomenon of spatio-temporal
chaos. Loosely speaking, this refers to systems that combine a familiar
temporal chaotic evolution with an additional decay of spatial correlations.
In a spatio-temporal chaotic regime both space translations and
time evolutions exhibit instabilities and it is even possible to define
spatial and temporal Lyapunov exponents\cite{Lepri:96}. One possible simple
mechanism for obtaining spatio-temporal chaotic motion is to spatially couple
low-dimensional chaotic units; although one has to be careful
because the coupling sometimes tends to reduce the spatio-temporal
instabilities\cite{Braiman:95}. However, it is also possible to produce
spatio-temporal chaos through the spatial interaction of well-behaved
(non-chaotic) units. This is the case for some metapopulation dynamics
models\cite{Octavio:95}.

In contrast with non-spatially distributed systems, spatio-temporal
systems possess a spatial extent. This may be discrete, giving a lattice
with a local dynamical unit at each site, or continuous. The typical model
in the latter case (if time is also continuous) is a partial differential
equation (PDE). By discretising space we obtain a lattice of ordinary
differential equations (commonly referred  to as a lattice differential
equation). In the discrete space case, the system can be viewed as a
collection
of low dimensional dynamical systems coupled together via some spatial rule
(for a review of current research see  Ref.~\cite{LatticeDynamics:97}).
Examples of this kind of model are widespread in the literature, particularly
in the field of solid state physics where they are used to study the
dynamics of interacting atoms arranged in a lattice (see
Ref.~\cite{Poggi:97} and references therein).

In this paper we focus our attention on a third category of
extended dynamical systems where not only space but also time is discrete.
In such a case the model consists of low-dimensional dynamical units with
discrete time ({\em i.e.}~maps) arranged in some discrete lattice
configuration in one or more spatial dimensions. Such models are usually
called coupled map lattices (CMLs). They were first introduced in 1984 as
simple models for spatio-temporal
complexity\cite{Kan:84,Waller:84,Crutchfield:84}. Despite their
computational simplicity, CMLs are able to reproduce a wide variety
of spatio-temporal behaviour, such as intermittency\cite{Keeler:86},
turbulence\cite{Beck:94,Kan:89} and pattern formation\cite{Kan:book},
to name just a few.

In this paper we shall focus on a one-dimensional  array of sites. The
length of this array could be infinite but here we restrict ourselves to an
array of length $N$ with periodic boundary conditions. At the $i$-th site
we introduce a discrete time local dynamical system whose state at time $n$
we denote by $x_i^n$. We suppose that the same local map $f$ acts at every
spatial location (so that the local dynamics is homogeneous). In the
simplest case, the local variable $x_i^n$ is taken to be one-dimensional.
The dynamics of the CML is then a combination of the local dynamics and the
coupling, which consists of a weighted sum over some spatial neighbourhood.
The time evolution of the $i$-th variable is thus given by
\begin{equation} \label{gral-CML}
x^{n+1}_i = \sum_k \e_k f(x^n_{i+k}),
\end{equation}
where the range of summation defines the neighbourhood.
The coupling parameters $\e_k$ are site-independent,
and satisfy $\sum \e_k=1$.
The commonest choice for the coupling scheme is
\begin{equation} \label{diffu}
x^{n+1}_i  =  (1-\e) f(x^n_i)
          \ds + {\e\over 2}\left( f(x^n_{i-1}) + f(x^n_{i+1}) \right),
\end{equation}
which is sometimes called a {\em diffusive\/} CML. This is a discrete
analogue of a reaction-diffusion equation. There is now a single coupling
parameter $\e$ which is constrained by the inequality $0\leq\e\leq1$, to
ensure that the signs of the coupling coefficients in (\ref{diffu})  
({\em i.e.}~$\e/2$ and $1-\e$) remain positive.

\section[LS for extended dynamical systems]
        {Lyapunov spectra for extended dynamical systems\label{LS-KNOWNJ}}

Let us now give an overview on the extraction and application of
Lyapunov spectra for spatio-temporal systems.
For a $N$-dimensional dynamical system there exist $N$ Lyapunov
exponents which in principle can be obtained from the
eigenvalues of the matrix
\begin{equation} \label{Gamma}
\Gamma=\lim_{n\rightarrow\infty}{\left[P(n)^{\rm tr}\cdot P(n)\right]^{1/2n}},
\end{equation}
where $P(n)$ corresponds to the product of the first $n$ Jacobians along
the orbit and $(\,\cdot\,)^{\rm tr}$ denotes matrix transpose. It is
well known that given an invariant measure (which is usually assumed to be
the natural measure for the dynamics) the Multiplicative Ergodic Theorem
({\em e.g.}~Ref.~\cite{Eckmann:85}) ensures that the limit exists for 
almost all initial conditions $x^0$, and if the measure is ergodic then its
eigenvalues are independent of $x^0$. The Lyapunov exponents are then
defined as the logarithms of these eigenvalues (which are obviously
non-negative). Although computing the exponents using equation (\ref{Gamma})
appears straightforward, in practice multiplying the Jacobians
is an ill-conditioned procedure since the most expanding
direction swamps all the other expansion/contraction rates.
Therefore one has to resort to algorithms that regularly reorthogonalise
the product of the Jacobians. Such algorithms perform a QR decomposition
every few iterates and are closely related to standard methods of computing
eigenvalues\cite{Geist:90}. The QR decomposition can be carried
out using Gram-Schmidt, Householder or Givens based techniques.
In this paper we use an efficient Householder method where only the terms
required for the computation of the Lyapunov exponents are actually
calculated\cite{Bremen:97}.

We define the Lyapunov spectrum (LS) as the set of Lyapunov
exponents $\{\lambda_i\}_{i=1}^N$ arranged in decreasing order.
The LS not only gives the expansion/contraction
rates of infinitesimal perturbations, but can also provide estimates of
fractal dimensions and entropies. Thus, for instance, the dimension of the
chaotic attractor ({\em i.e.}~informally the effective number of degrees of
freedom) is given by the Kaplan-Yorke conjecture\cite{KY:conjecture}
through the {\em Lyapunov dimension}
\begin{equation} \label{Lyapunov_dimension}
D_L=j+ {1\over | \lambda_{j+1} |} \sum_{i=1}^{j} \lambda_i,
\end{equation}
where $j$ is the largest integer for which $\sum_{i=1}^{j} \lambda_i>0$.
It is also possible to extract an upper bound for the
{\em Kolmogorov-Sinai {\rm (KS)} entropy} $h$ from the LS by the following
approximation\cite{Eckmann:85}
\begin{equation} \label{entropy}
h= \sum \lambda_i^+,
\end{equation}
where the summation is over the positive Lyapunov exponents $\lambda_i^+$.
The KS entropy quantifies the mean rate of information production in a
system, or alternatively the mean rate of growth of uncertainty due to
infinitesimal perturbations.

As an example, figure \ref{cmllogLSF.ps}.a shows the LS for a coupled
logistic lattice of size $N=20$ with periodic boundary conditions obtained
using the known dynamics (\ref{diffu}). In
figure \ref{cmllogLSF.ps}.b we plot the sum of the
Lyapunov exponents $\sum_{k=1}^{i} \lambda_k$ from where
the KS entropy can be estimated (maximum of the curve)
and the Lyapunov dimension can be extracted (intersection
with the horizontal axis). Notice that the
Lyapunov dimension $D_L\simeq 15.4$ is comparable to
the total size of the lattice $N=20$.
Also observe that $D_L\simeq 15.4$ implies the presence of a high
dimensional attractor. This point will be addressed in more detail
in the following sections.

The computation of the whole LS for spatio-temporal systems is a cumbersome
task, particularly as the size of the system increases. Even the most
efficient methods involve ${\mathcal O}(N^3)$ arithmetic operations per
time step\cite{Geist:90} for a lattice of size $N$. Even for moderate
lattice sizes ($N\sim 50$) the number of operations required is already
considerable.
Moreover, taking into account that sometimes the (time) convergence of the
Lyapunov exponents is extremely slow, the time necessary to compute
the whole LS for spatio-temporal systems quickly becomes prohibitive.
Furthermore, one is often interested in the behaviour of the system in the
thermodynamic limit where the number of lattice sites $N$ goes to
infinity. In such a case it quickly becomes impossible to compute the whole
LS and one has to resort to subsystem rescaling techniques that we
shall now briefly describe.

\twoFIG{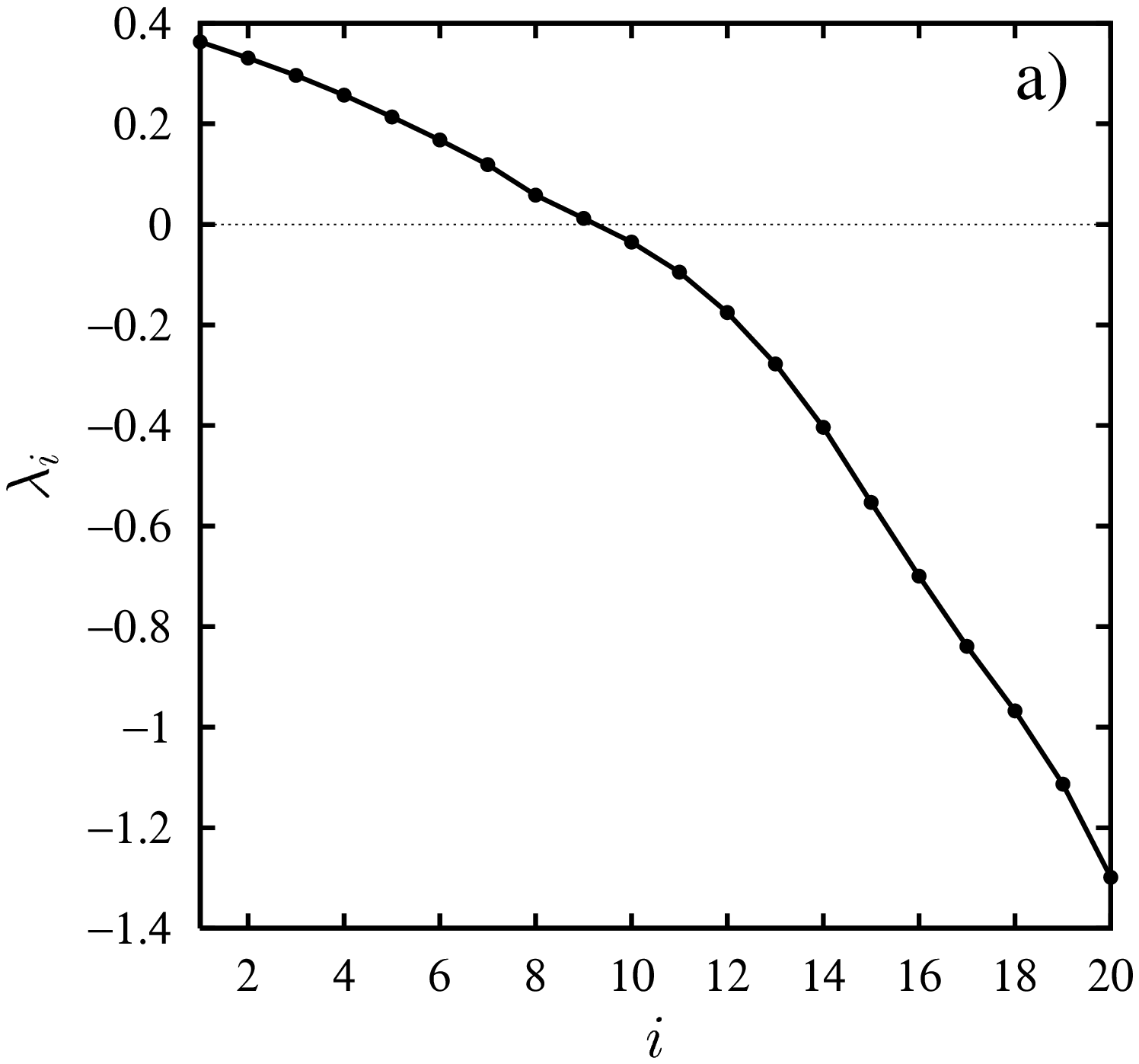}{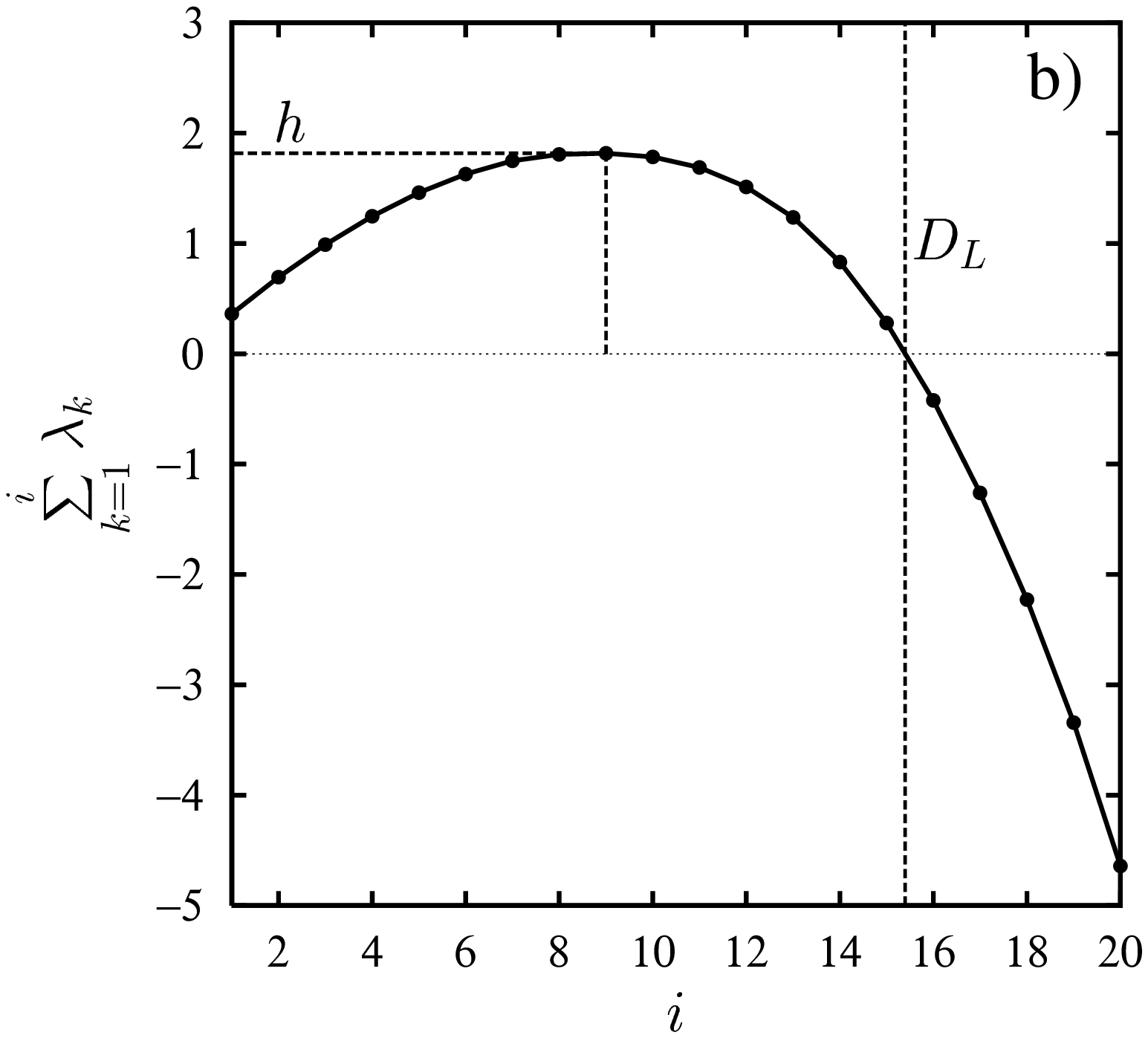}{\FigSize}{\CMLLOGLSCAP}

Consider a one-dimensional lattice of coupled one-dimensional dynamical
units. High dimensional cases may be treated in a similar way. Assume
the array is large ($N\gg 1$) and suppose initially that it is made up of
smaller {\em independent} (uncoupled) subsystems of size $N_s$. 
For simplicity take $N$
to be a multiple of $N_s$. The LS of the whole lattice then consists of
$N/N_s$
exact copies (assuming spatial homogeneity) of the subsystem LS
of $N_s$ Lyapunov exponents. Thus, the LS for the whole lattice consists of
$N_s$ Lyapunov exponents, each with multiplicity $N/N_s$. As we relax the
assumption that the subsystems are uncoupled we hope that the LS does not
change significantly\cite{Ruelle:82,Ruelle:83}. If this is the case, the
whole LS of the original lattice may be approximated by a rescaled version
of its subsystem LS.

An obvious choice for such a rescaling is to multiply the index $i$
of the $N_s$-subsystem Lyapunov exponents $\lambda_i(N_s)$ by
$r'=N/N_s$. However, careful analysis of the LS of a homogeneous state
suggests that a better rescaling for one-dimensional lattices
is given by\cite{rcg:sublya}
\begin{equation}\label{rescaling}
r={N+1 \over N_s+1}.
\end{equation}

The rescaling of the whole LS from subsystem information leads us
naturally to define intensive quantities from extensive quantities. As a
direct consequence of this rescaling, using the same argument as in the
previous paragraph, it is straightforward to see that the Lyapunov
dimension and the KS entropy are extensive quantities ({\em i.e.}~$D_L$
and $h$ increase linearly with $N$). Moreover, it is useful to introduce
their respective densities by simply dividing them by the system volume
(lattice size)\cite{Parekh:98}. This even allows such quantities to be
defined in the thermodynamic limit, although care has to be taken in
doing this for systems such as PDEs where space is continuous\cite{Collet:98}.

By using subsystem rescaling one can considerably
reduce the computing resources required
to estimate the whole LS of a large spatio-temporal system.
Furthermore, by restricting oneself to observing data from a subsystem,
one reduces the dimensionality of the resulting time series and increases the
likelihood of success in applying reconstruction techniques.

\section[LS from univariate time series]
        {Lyapunov spectra from univariate time series \label{LS-UNIVARIATE}}

In the previous section we assumed that the dynamics of the system whose LS
we wished to compute was known. This is not the case for many complex
physical systems. We therefore now turn to the task of estimating the LS
when the only information available about the system is a time
series of observed data. Let us start by applying standard delay
reconstruction techniques for
a single observable time-series. Thus suppose that the univariate time series
$\{\varphi^n\}$ has been measured from a $k$ dimensional system. We can
create a high dimensional reconstructed state space by constructing the
delay vectors
\begin{equation}\label{T-delay}
y^n=(\varphi^n,\varphi^{n-1},\dots,\varphi^{n-(d-1)}),
\end{equation}
where $d$ is the so-called {\em embedding dimension}. Takens' 
Theorem\cite{Takens:81} says that for $d\ge 2k+1$, the dynamics 
induced on such delay vectors is generically smoothly conjugate 
to the original dynamics.  
This apparently arbitrary condition is a simple geometrical 
disentangling in order to avoid self intersections on the 
reconstructed manifold\cite{Whitney:36}. 
The smooth conjugacy between induced and original dynamics
asserts that the time series contains all the coordinate-free
properties of a dynamical system. Specifically, since the LS is
invariant under smooth conjugacy ({\em i.e.}~smooth coordinate changes)
the LS computed from the dynamics of the delay vectors will be the same
as that of the original system. Note that the time series need not to be
the time history of one of the system's state variables but can be any
generic function of the state of the system. In particular,
in principle it is possible to reconstruct all the Lyapunov
exponents from a univariate time series. To anyone familiar with the
proof of Takens' Theorem is it obvious that it generalises
straightforwardly to multi-variate time series, though as far as we are
aware, a proof has never appeared in the literature.

Unfortunately, Takens' Theorem does not actually apply to the CMLs
considered here\cite{Sakse:98}. Firstly, the maps that we use as local
dynamics are not invertible, and the proof of Takens' Theorem depends
fundamentally on such invertibility. Secondly, we assume translation
invariance of the CML ({\em i.e.}~the dynamics and coupling are
spatially homogeneous) and such symmetry renders our systems far from
generic. Finally, in a similar fashion, measurement functions depending
on only a single site (or group of sites) are not generic in the space
of all observables, indeed a generic function will depend on {\em all}
$N$ sites. On the other hand our results, and in particular the
comparison with the LS computed in the previous section assuming full
knowledge of the actual dynamics, suggest that we are able to estimate
the real LS from observed time series. We shall therefore use delay
reconstruction techniques throughout this paper without worrying about
this failure of Takens' Theorem (and in any case the theorem only draws
conclusions for generic dynamics and measurement functions, and so cannot
guarantee reconstruction for any particular time series).

For the extended dynamical systems we are considering, the dimension $k$ is
the size of the lattice $N$ times the dimension of the local dynamics.
Hence for the logistic coupled map lattice in figure \ref{cmllogLSF.ps} we
get $k=N$.
Even for quite moderate $N$ an embedding dimension of $2N+1$ is clearly
quite impractical. Of course, the condition $d \ge 2k+1$ is only a
sufficient one, and it is well known that some systems ({\em e.g.}~the
Lorenz equations) can be reconstructed for smaller values of $d$.
Furthermore, Sauer and Yorke\cite{Sauer:99} have a sharper estimate for
$d$ that is still sufficient to preserve the box-counting dimension and the
Lyapunov exponents
\begin{equation}\label{D_Sauer}
d > D_B + D_T,
\end{equation}
where $D_B$ is the box-counting dimension of the attractor and $D_T$ is a
``tangent dimension'' which is informally the maximum (over all points)
dimension of the tangent space of the underlying dynamics. However even
this is still very prohibitive. For example, for the system in figure
\ref{cmllogLSF.ps}) the Lyapunov dimension for even quite a small lattice
of $N=20$ coupled logistic maps we have $D_L \sim 15.4$. Then assuming 
$D_B=D_L$ ({\em i.e.}~essentially the original Kaplan-Yorke conjecture),
the bound (\ref{D_Sauer}) gives $d > 15.4 + D_T$, and one would typically
expect $D_T$ to lie between $D_B$ and $N$, giving a value of $d$ from 30
upwards. Of course the Sauer-Yorke inequality is also just a sufficient
condition, but if one is to preserve all $N$ exponents, it seems that an
absolute minimum of $d \ge N$ will be necessary, and even this is too large
to be generally practicable.

\oneFIG{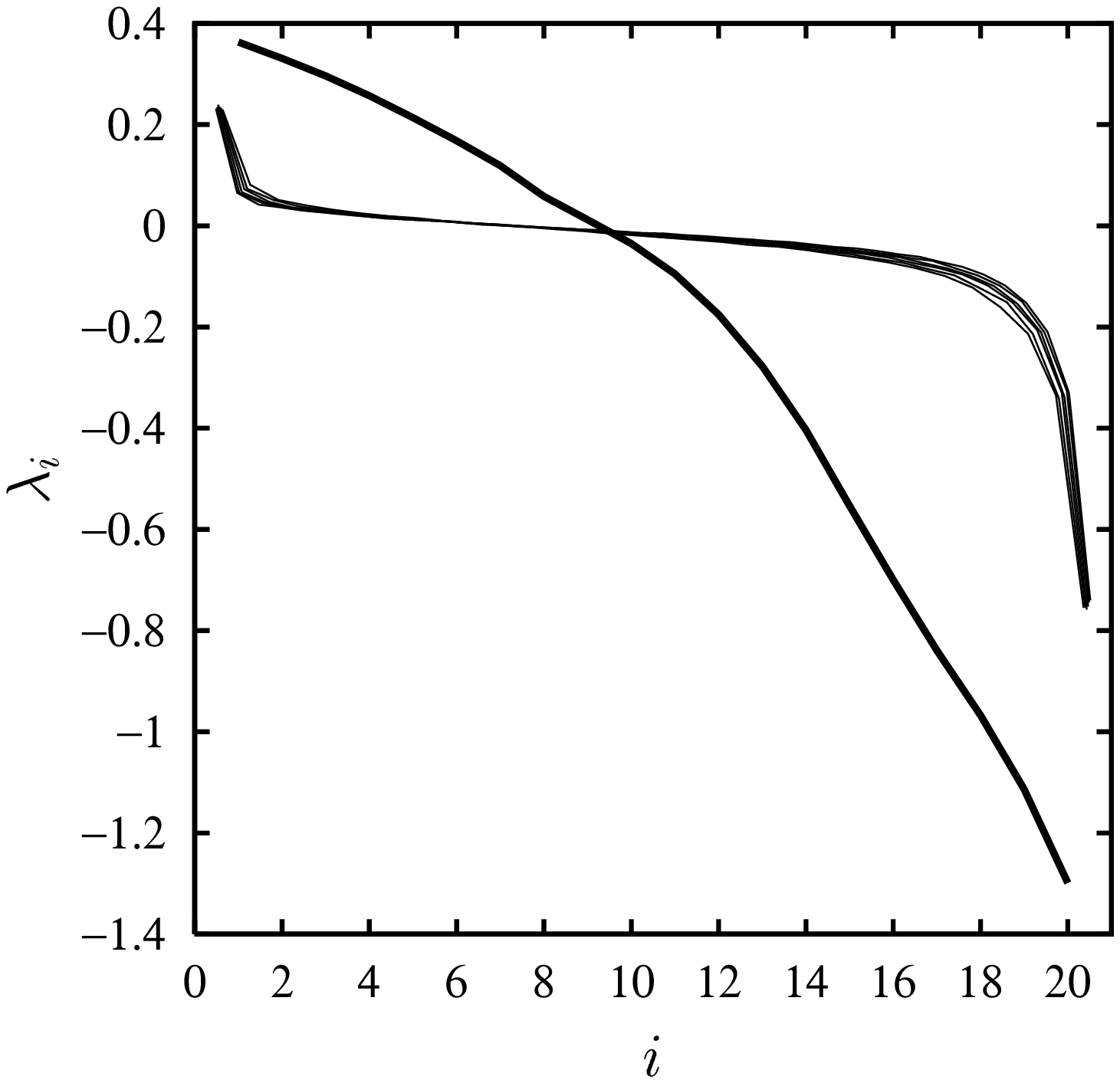}{\FigSize}{\CMLLOGDSONECAP}

Indeed, as indicated in the introduction, an embedding dimension much
larger than $d\sim 6$ will lead to the so-called ``curse of
dimensionality''. Thus in order to estimate the LS from the time series we
need to approximate the map that governs the dynamics of the delay vectors,
that is the map $F$ such that $F(y^n)=y^{n+1}$. There exist several methods
for doing this, of which the best for the purpose at hand are all based on
local approximations. These can be constructed by scanning the time series
to find
a set of past close encounters (corresponding to neighbours in the
reconstructed state space) of the current reconstructed
state $y^n$. Once these neighbours have been found, an approximation
$\hat F$
to $F$ at $y^n$ can be computed using, for example, a
least squares fit within an appropriate space of functions ({\em e.g.}~linear 
or quadratic). For a comprehensive review see the recent book by
Kantz and Schreiber\cite{Schreiber:book}. For large embedding dimensions
($d\sim 6$) it becomes extremely difficult to find close neighbours and
hence, due to data
limitations, the approximation to $F$ ceases to be local. This in turn can
lead to large errors in the determination of the Lyapunov exponents.

It is clear then that function approximation techniques are bound to fail
when reconstructing spatio-temporal systems where the dynamics
is typically high dimensional. As an example figure
\ref{cmllogDS1F.ps} shows the results of an attempt to estimate the whole LS
from a univariate time series obtained from a single site
of a couple logistic lattice. The measurement function in this case
is the projection onto a single site, so that
$\varphi^n=x_j^n$ for a fixed $j$. An embedding dimension in the range
$32\le d \le 42$ was used. It is obvious from the figure that this temporal
delay reconstruction of the LS (overlapping thin lines) completely fails
to reproduce the LS of the original system (thick line).
Other values of the reconstruction parameters than the ones given
in the figure caption were also tried yielding the same
qualitative results.

\section[LS from spatio-temporal data]
        {Lyapunov spectra from spatio-temporal data \label{LS-MULTIVARIATE}}

Since the direct application of univariate (temporal) delay
reconstructions clearly fails for spatio-temporal systems,
let us now try to exploit the spatial extent of the system and
use a spatio-temporal delay reconstruction. The framework for this is
presented in\cite{Sakse:98}. Here, we shall show the benefits of including
spatial information in estimating the LS. We shall also investigate the
advantages of truncating the outer layer of a reconstructed Jacobian in
order to avoid boundary effects. We have already addressed these issues in
a previous paper\cite{Sakse-rcg:99}, however we repeat some of the analysis 
here for completeness.

A spatio-temporal reconstruction is obtained by replacing the delay vectors
(\ref{T-delay}) of the previous section by the {\em spatio-temporal} delay
vectors
\begin{equation}\label{ST-delay}
y^n_i=(\phi^n_i,\phi^n_{i-1},\dots,\phi^n_{i-(d_s-1)}),
\end{equation}
whose entries $\phi^n_i=(x_i^n,x_i^{n-1},\dots,x_i^{n-(d_t-1)})$ are
time-delay vectors and the spatial index $i$ is fixed. The overall embedding
dimension for such a spatio-temporal reconstruction is $d=d_s d_t$, where
$d_s$ and $d_t$ denote the spatial and temporal embedding dimensions
respectively. The standard temporal delay reconstruction (\ref{T-delay})
is recovered by setting $d_s=1$. Spatio-temporal reconstructions have
proved useful in various contexts\cite{Muldoon:94,Cao-Mees-Judd:98,Leung:95}.
By adding the extra spatial delay in the time series one
is including important information that is otherwise very difficult
to obtain. In fact, it is extremely difficult to extract information
about neighbouring dynamics just by doing spatially localised
measurements. This is because of the rapid decay of spatial correlations
in locally coupled extended dynamical systems. In a previous 
paper\cite{rcg:trunca} we showed how this extremely rapid decay in
correlations allows a quite small truncated lattice with random
inputs at the boundaries to reproduce the local dynamics of a large,
potentially infinite, lattice. This tends to suggest that a univariate
time series cannot in practice reconstruct the dynamics
of a whole spatio-temporal system.

\oneFIG{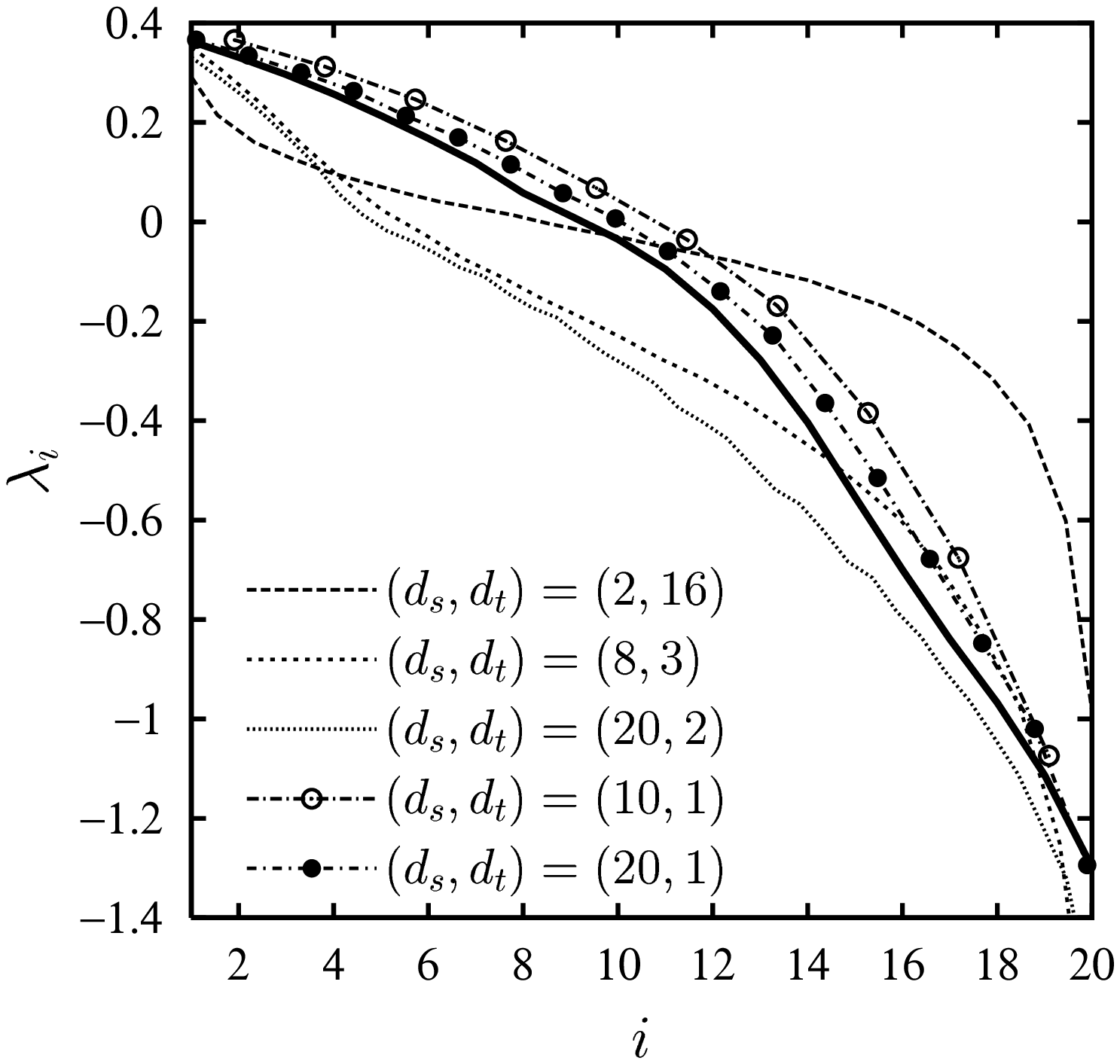}{\FigSize}{\CMLLOGDSDTCAP}

In figure \ref{cmllogDSDTF.ps} we present several estimates of
the LS of the coupled logistic lattice (as in figure \ref{cmllogDS1F.ps})
for different spatio-temporal embedding dimensions $d$. The different
reconstructions correspond to a broad sample of choices of $d_s$ and $d_t$.
Note that simply increasing $d=d_s d_t$ does not improve the estimates. In
fact, in order to get a good estimate of the LS, it seems preferable to
take $d_t=1$ and $d_s$ as large as possible. This is indicated by the empty
and filled circles corresponding to $d_t=1$ and $d_s=10$ and $d_s=20$
respectively. In the figure we show only a few combinations of $(d_s,d_t)$.
However, we originally considered a much broader choice without finding any
qualitative differences.
We therefore conclude that a spatio-temporal reconstruction with $d_t=1$
(that is a pure {\em spatial} delay reconstruction) gives the best results.
This is easy to explain since we are using the actual dynamical
variables of the system ($\varphi_i^n=x_i^n$) as observables and are
measuring them in a complete window. Thus, we do not need to increase
the embedding dimension to avoid self intersection in the reconstructed
space since we are already in a natural space for the system.

\oneFIG{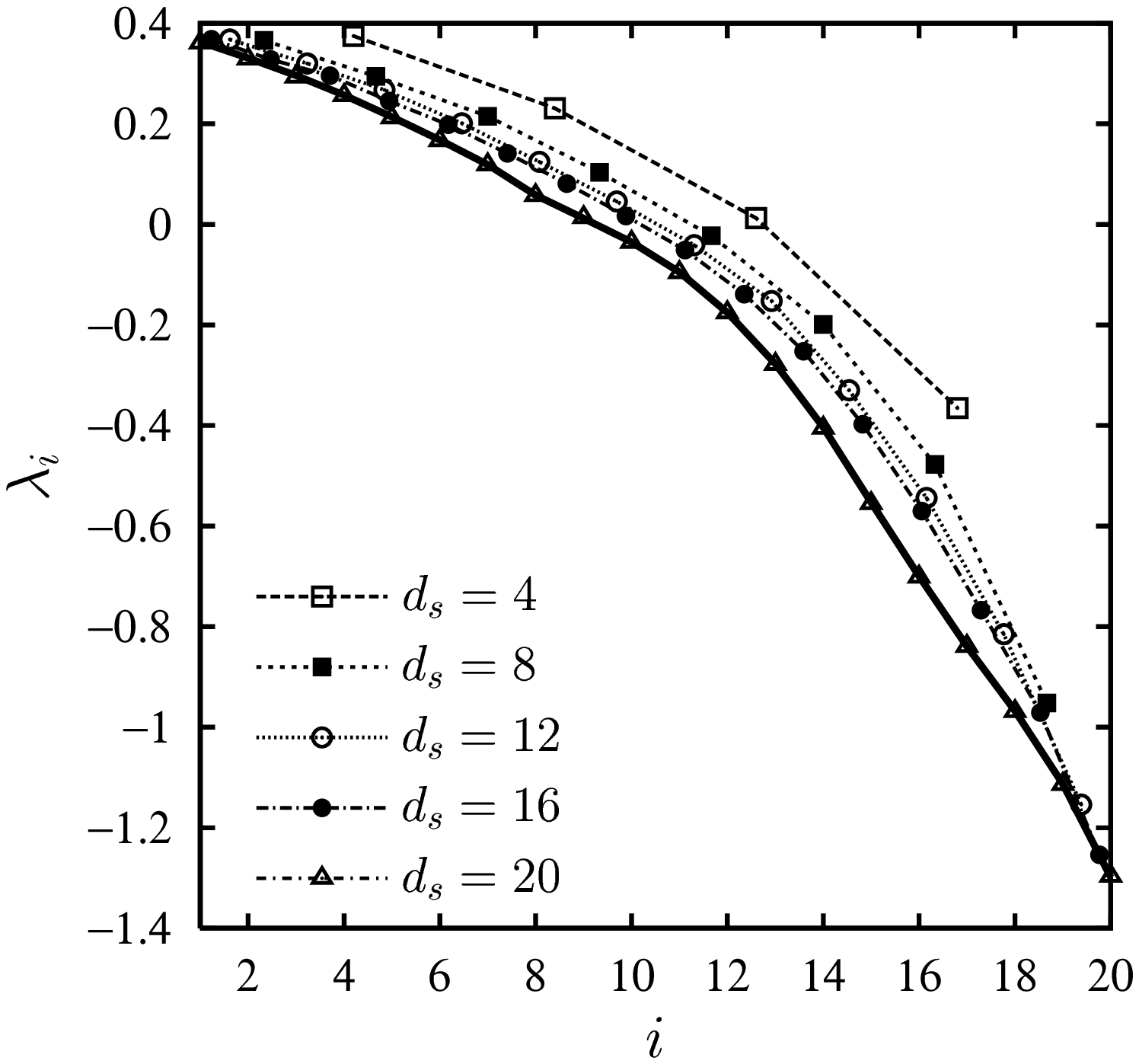}{\FigSize}{\CMLLOGDTONECAP}

In figure \ref{cmllogDT1F.ps} we show the reconstruction of the LS
using a pure spatial reconstruction with increasing $d_s$. These estimates
are much more promising than before. Note that since our
original system is of size $N=20$ it is not possible to choose
$d_s>20$. Indeed, we have to be careful, since
taking $d_t=1$ and $d_s=N=20$ corresponds to measuring the
{\em whole} state space and is therefore not a genuine delay reconstruction.
This explains why the reconstructed LS for
$d_t=1$ and $d_s=N=20$  agrees so well with the direct computations
(triangles in figure \ref{cmllogDT1F.ps}) though note that in figure
\ref{cmllogLSF.ps} we know the Jacobian of the dynamics exactly, whilst
here we still have to do some function fitting.

From figure \ref{cmllogDT1F.ps} we see that as $d_s$ is increased
a better estimate of the LS is obtained. One may think that
this is simply the effect of getting a higher embedding dimension.
However, the reason for the apparent convergence of the reconstructed
LS towards the exact LS as $d_s\rightarrow N$ is simply due to a
reduction of the noise from boundary effects.
Consider the sites at the boundary of our spatial window in which we
measure the multivariate time series. The dynamics of these sites
depends on sites whose evolution is {\em not} explicitly contained
in the measurement window. As a consequence, the estimates of the
Jacobian for these sites contain spurious terms and are
subject to error. For instance, the fitting procedure for the first row
tries to compensate for the lack of information from the left neighbouring
site
that is outside of the measurement window by including an artificial
dependence on {\em all} sites inside the window. This will yield erroneous
non-zero terms off the tri-diagonal.

\oneFIG{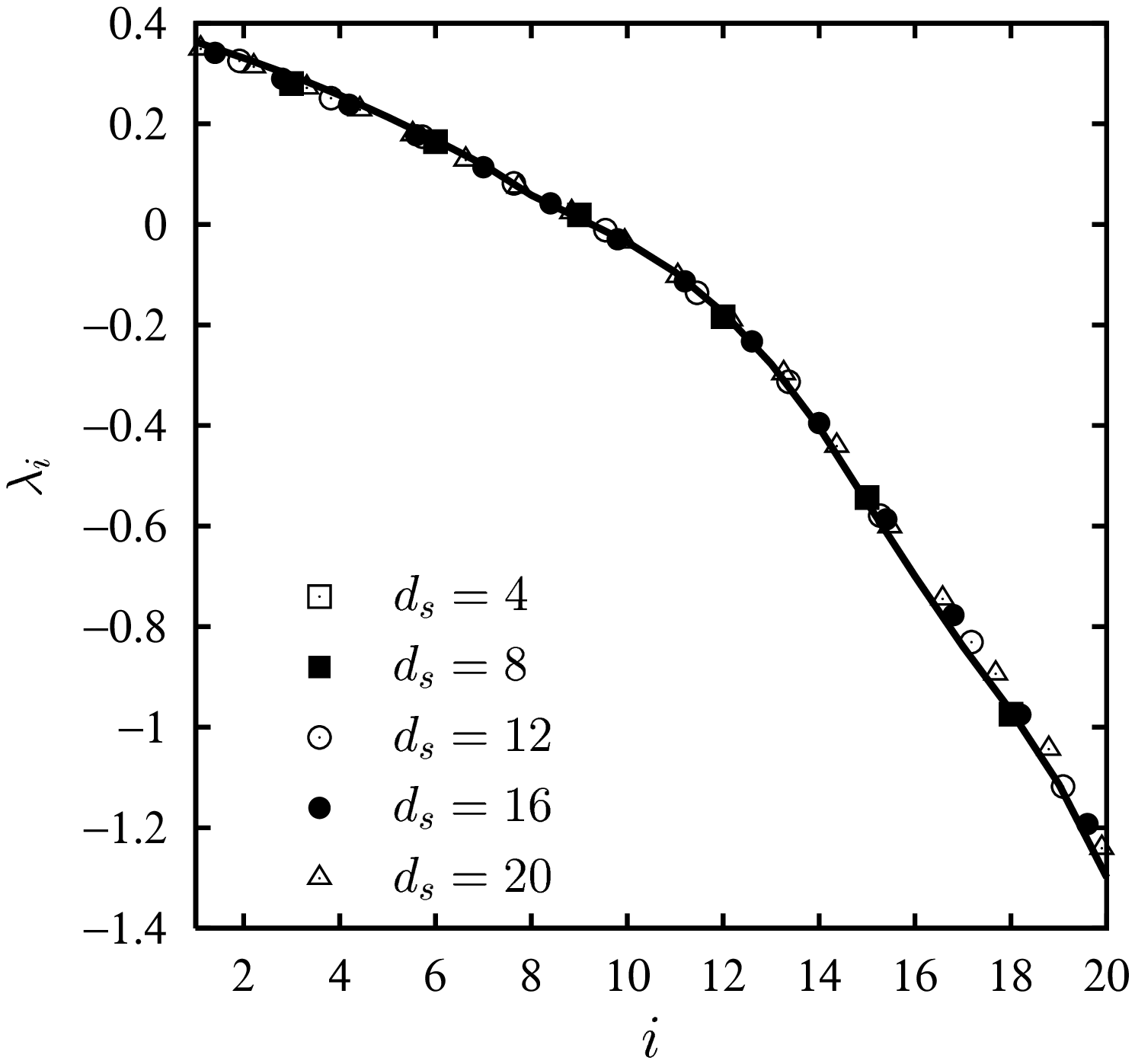}{\FigSize}{\CMLLOGDTONECONECAP}

We can avoid these effects by observing that the boundary sites
contribute towards the Jacobian at its outer rows and columns. We can
therefore remove such boundary effects by simply truncating the outer
rows and columns of the reconstructed Jacobian to yield a
$(d_s-2)\times(d_s-2)$ matrix\cite{Sakse-rcg:99}. Note that one must
first fit the $d_s\times d_s$ Jacobian and only then truncate its outer
rows and columns. Also observe that for a spatial delay embedding
dimension $d_s$ we obtain $d_s-2$ Lyapunov exponents. The results of
applying this technique are shown in figure \ref{cmllogDT1C1F.ps}.
Comparing this to figure \ref{cmllogDT1F.ps} where the outer layer of
the Jacobian was not truncated we see a dramatic improvement, and in
particular can obtain good estimates even for quite small $d_s$ (see
also Ref.~\cite{Sakse-rcg:99}). It is obvious that the truncation of the
outer layer of the Jacobian has to coincide with the range of the
coupling: the larger the coupling range the more outer rows and columns
must be truncated. This is considered further below in section
\ref{LS-GRALCOUPLING}.

\section{Local versus global fitting: a matter of dimensionality
\label{LS-BASIS}}

It is somewhat surprising that the estimated LS
in figure \ref{cmllogDT1C1F.ps} are in such good agreement with the actual LS
despite the fact that the embedding dimensions are a) too small
to disentangle the state space and b) too large to avoid the
``curse of dimensionality''. The explanation for this apparent
contradiction is that the measurement functions
are the actual variables of the system and the form of the dynamical
equations is actually contained in the space spanned by the basis functions
used to approximate them. In particular, the dynamics of our CML is
quadratic in nature ({\em c.f.}~equation (\ref{diffu}) with 
logistic local maps) and we are using a quadratic fit.
Thus, a local fit is in fact also a global one and the problem of not
finding enough neighbours in high-dimensions is avoided. It thus turns out
to be possible to obtain very good estimates of the dynamics with even
moderate amounts of data.

\oneFIG{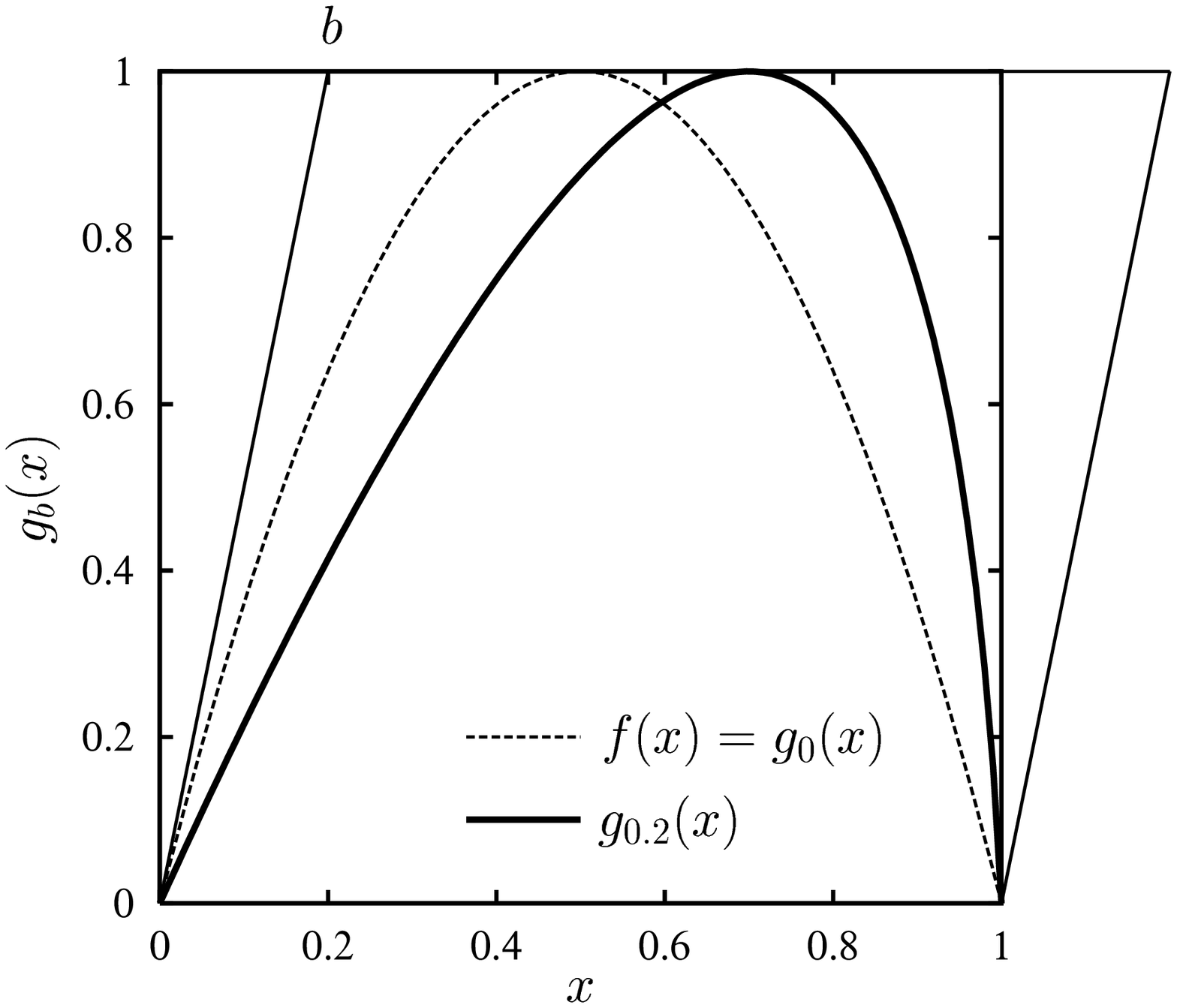}{\FigSize}{\SKEWCAP}

This situation is however unlikely to arise in most real applications, and
typically we expect that the (delay reconstructed) dynamics will not lie in
the space spanned by the basis functions. To investigate the effects of
this we now turn our attention to a skewed logistic local map which yields
a CML whose dynamics is not contained in the space of quadratic polynomial
functions that we use for fitting. This map is obtained from the standard
logistic map $f(x)=ax(1-x)$ by applying the following skew transformation
of the unit square
\begin{equation}\label{skew-transform}
K(x,y) = (x+by,y),
\end{equation}
where $b$ is a parameter determining the degree of skew. In this way we obtain
the map (figure \ref{skewF.ps})
\begin{equation}\label{skew}
g_b(x) = {-1+ab(2x-1) + \sqrt{(1+ab)^2-4abx} \over 2ab^2},
\end{equation}
where $b$ must satisfy $-1/a<b<1/a$ so that the derivatives of $g_b$ at 0
and 1 remain bounded.

\oneFIG{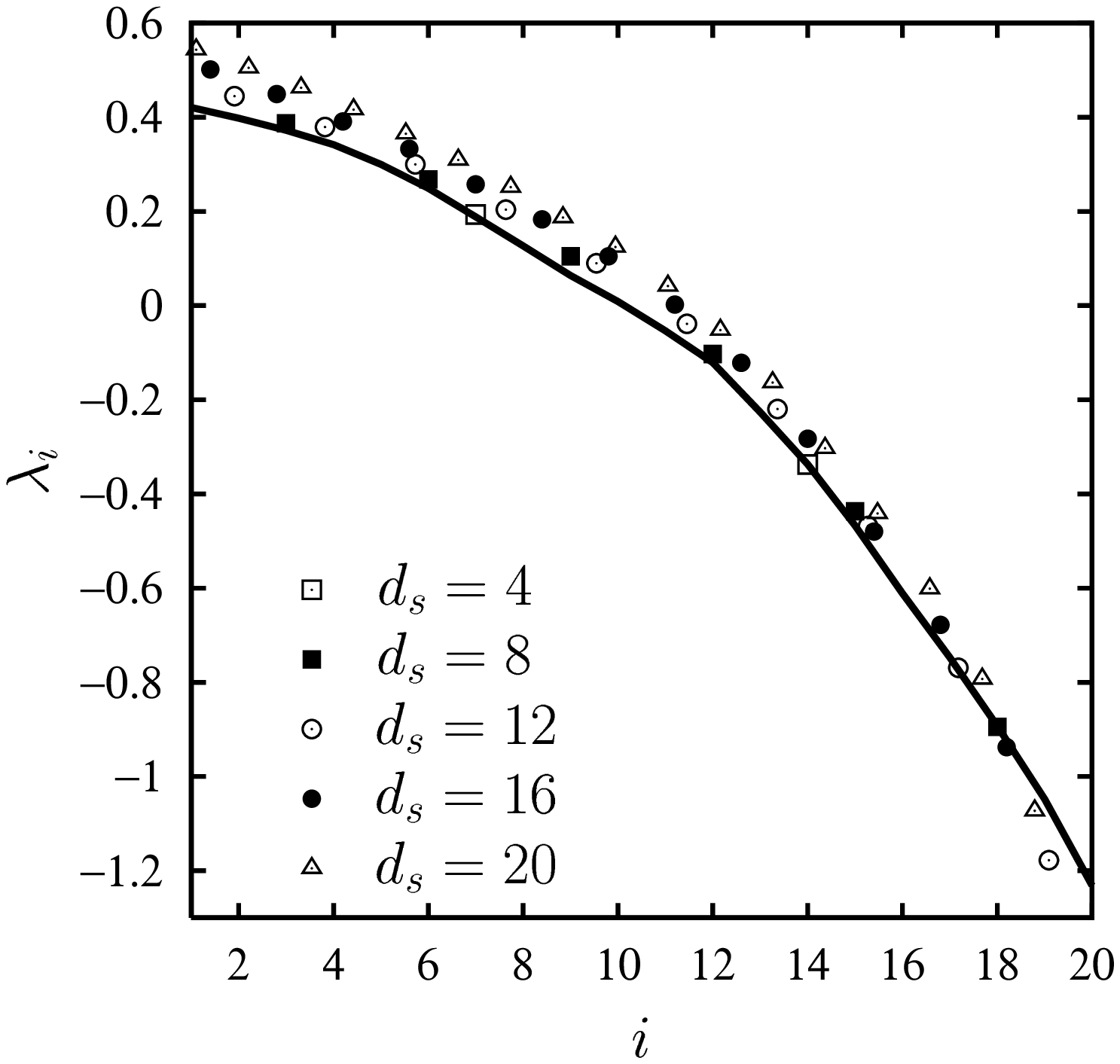}{\FigSize}{\SKEWBTWOCAP}

We now repeat the estimation of the LS for this skewed map. Figure
\ref{skewDT1b2F.ps} shows the effects of increasing the spatial embedding
dimension $d_s$ in the case of large skew $b=0.2$. Observe that for small
$d_s$ ($\leq 6$) the approximation of the LS is quite good
(see square points in figure). However, as
$d_s$ increases, the estimate deteriorates. This reflects the fact that as
the embedding dimension
grows the fitting algorithm has more and more difficulty finding close
neighbours. Since the skewed logistic map cannot be well
approximated globally by a quadratic function the LS estimate deteriorates.
Note however that even though an accurate estimate of the LS
is not possible in this case, the estimated LS does have the right
qualitative features. This is due to the fact that the skewed logistic map
still resembles
a quadratic function to some extent and thus a global fit is still
reasonable. This is illustrated in figure \ref{skewDT1s20F.ps} where we
present the effect of varying the degree of skew $b$, and hence the degree
of departure from the space of quadratic functions. We see that as the skew
is increased, the error between the actual and the estimated LS rapidly
increases. Using a local map which is even further from the space of
quadratic ones leads to yet greater discrepancies\cite{Sakse-rcg:99}.

Finally recall also that the number of Lyapunov exponents obtained for a
given $d_s$ is $d_s-2$ due to the truncation of the outer layers of the
Jacobian. Therefore if we try an avoid the ``curse of dimensionality'' by
working in low embedding dimensions with $d_s\leq 6$ we obtain only 4
points or less on the LS density curve. Even if these are accurate, we
cannot in general expect to extrapolate the whole spectrum from just these
4 points. Hence estimates of Lyapunov dimension and KS entropy are likely
to be quite poor\cite{Sakse-rcg:99}.

\oneFIG{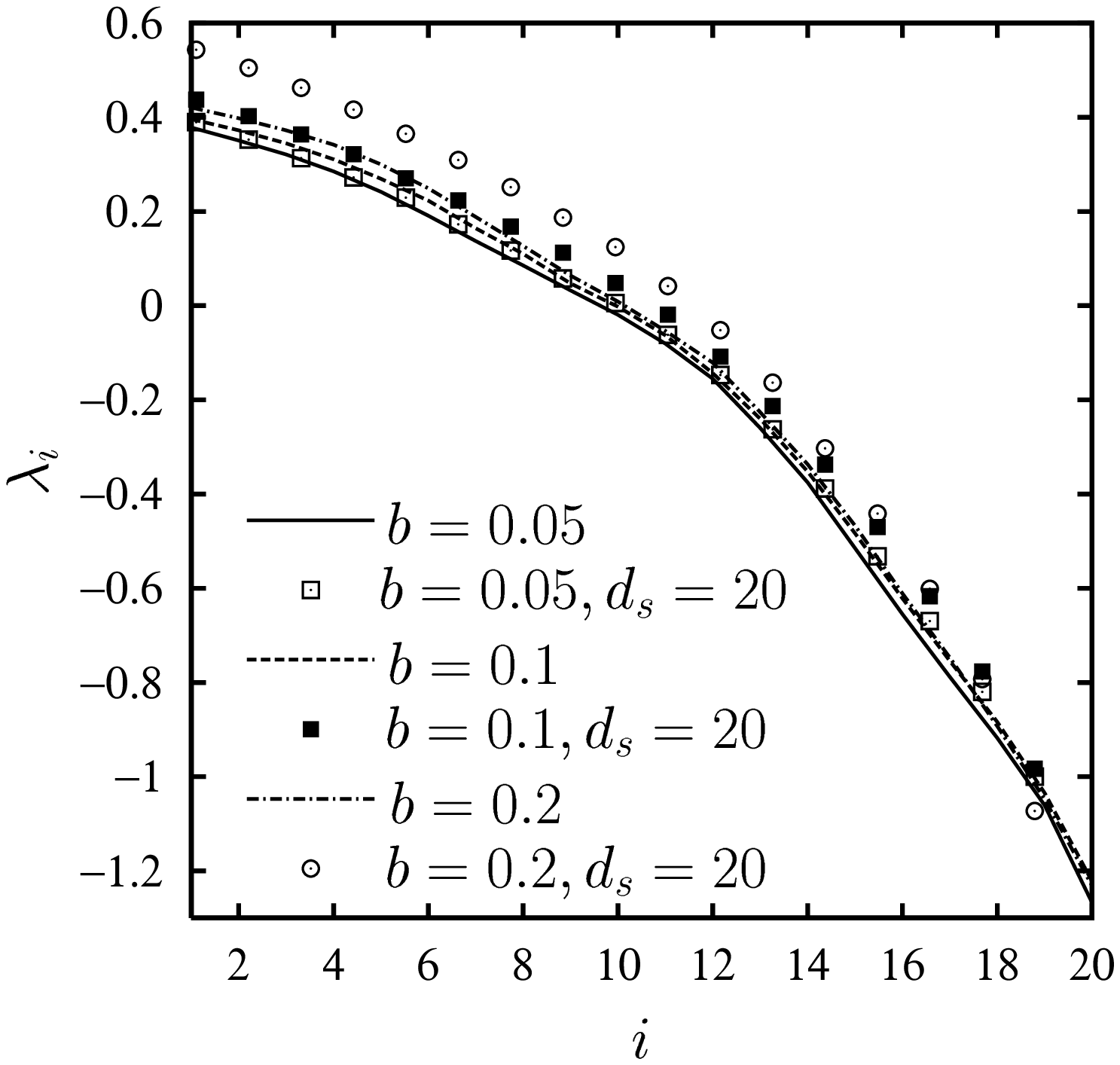}{\FigSize}{\SKEWSTWENTYCAP}

The results presented in this section suggest that as soon as the
dynamics cannot be globally approximated by the space spanned by the
basis functions used in the fit, and as soon as more than a handful of
Lyapunov exponents is required, standard spatio-temporal embedding
techniques cannot reliably estimate the LS. In the next section we shall
show how this difficulty can be overcome by focusing on the estimation
of only the nontrivial entries in the Jacobian. An alternative approach
is to try and find an adapted fitting basis that will allow us to fit
the dynamics globally. This can be done by first extracting an estimate
of the local dynamics from the time series and then using this to
construct an appropriate set of basis functions. One possible way of
estimating the local dynamics is to use time-delay plots\cite{Sakse-rcg:99}. 
An even more promising technique is to consider quasi-homogeneous states 
in a small window and their time evolution\cite{rcg:loc-dyn}.

\section{Quasi-diagonal reconstruction of the Jacobian\label{LS-QUASIJ}}

In this section we improve the method presented above by making use of the
local nature of the coupling in our CML, which results in the Jacobian
exhibiting a {\em banded-diagonal structure}. It is thus natural to only
attempt to estimate the non-zero entries in this Jacobian. We call this a
{\em quasi-diagonal} reconstruction. This allows us to carry out the local
fit in a low dimensional space, and avoids the difficulties described
above. A similar idea was employed by B\"unner and 
Hegger\cite{bunner-hegger:99}. However, they only applied it to a standard
logistic CML, which as we have seen above is a relatively easy system for
which to estimate the spectrum using conventional means. It is thus
impossible to judge from their work how much of a benefit the
quasi-diagonal approach actually brings. Additionally, here we combine this
technique with a truncation of the outer layers of the Jacobian, which, as
we have seen above, brings substantial benefits, but was not considered 
in\cite{bunner-hegger:99}.

The idea behind a quasi-diagonal reconstruction of the Jacobian
is quite simple. The entries of the Jacobian $J(n)$ at time $n$ are given by
\begin{equation}\label{J_kl}
J_{kl}(n)={\p x^{n+1}_{k}\over\p x^{n}_{l}}.
\end{equation}
Due to our assumption of localised coupling most of these terms are
identically zero. The only non-zero terms lie within a distance of the
diagonal given by the distance over which the coupling acts. In the
particular case of a nearest neighbour CML (\ref{diffu}), the Jacobian is
tri-diagonal, and has only three non-zero elements in each row:
\begin{equation}\label{Jtridiag}
J(n)=\pmatrix{
%
\p^{1}_{1}&\p^{1}_{2}&    0     &      &\phantom{\ddots}&0\cr
\p^{2}_{1}&\p^{2}_{2}&\p^{2}_{3}&  0   &            &\phantom{\ddots}\cr
    0     &  \ddots  &  \ddots  &\ddots&      0     &            \cr
          &    0     &  \ddots  &\ddots&   \ddots   &    0       \cr
          &          &     0    &\ddots&   \ddots   &\p^{d_s-1}_{d_s}\cr
0&\phantom{\ddots}   &          &  0   &\p^{d_s}_{d_s-1}&\p^{d_s}_{d_s}~~~\cr
},
\end{equation}
where, for simplicity, we use the notation:
$$\p^{u}_{v}=\pp{i_0+u-1}{i_0+v-1}$$
Note that the Jacobian (\ref{Jtridiag}) is extracted from sites $i_0$ to
$i_0+d_s-1$ and so is of size $d_s\times d_s$. The starting index $i_0$
is arbitrary since we are assuming spatial homogeneity.

We estimate this Jacobian in a row by row fashion by fitting local dynamics
of the form
\begin{equation}\label{F_r}
x_{r}^{n+1}=F_r(x_{r-1}^n,x_{r}^n,x_{r+1}^n).
\end{equation}
for $(1<r<d_s)$. This fit is carried out in just a three dimensional space,
as opposed to a $d_s$ dimensional as before, and close neighbours are
defined by their distance from $(x_{r-1}^n,x_{r}^n,x_{r+1}^n)$. This thus
avoids any problems with high-dimensionality. Furthermore, if the dynamics
is translation invariant ({\em i.e.}~spatially homogeneous), we can use any
triple of the form $(x_{j-1}^n,x_{j}^n,x_{j+1}^n)$, regardless of position.
This dramatically increases the effective amount of data available to us
for performing the fit. Additionally, if the dynamics is also isotropic,
{\em i.e.}~invariant under the symmetry that exchanges $x_{j}^n$ with
$x_{N-j}^n$, we can use all triples in reverse order
$(x_{j+1}^n,x_{j}^n,x_{j-1}^n)$. Of course, if the coupling acts over a
longer range then the map $F_r$ will depend on more variables, but as long
as the coupling remains reasonably local this approach will still bring
advantages (we investigate this further in the next section).

\oneFIG{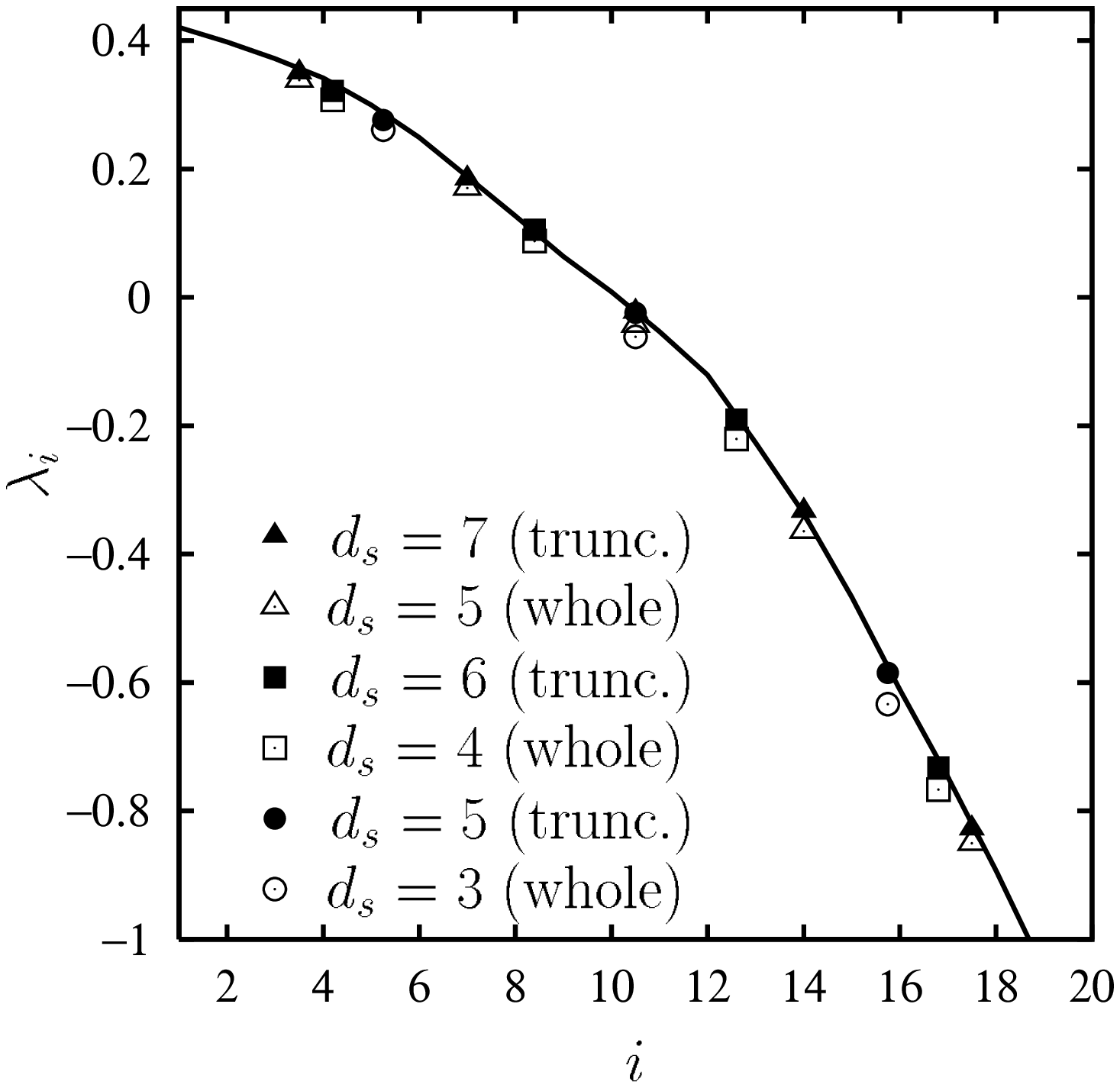}{\FigSize}{\SKEWNTCAP}

The first and last rows of $J(n)$ contain only two non-zero entries due to the
lack of information from outside the observation window. As before, the
entries will be estimated incorrectly since the fitting algorithm will
try and compensate for the lack of the missing entry by erroneously
adjusting the remaining two. We therefore truncate the outer layers of
the Jacobian as in the previous section and compute only $d_s - 2$
Lyapunov exponents from a window of size $d_s$. The improvements due to
such truncation are shown in figure \ref{skewNTF.ps}. Note that these
become less apparent as the width $d_s$ of the observation window
increases. This is consistent with the fact that the boundary effects
become less significant as the size of the subsystem grows\cite{rcg:trunca}. 
Nevertheless, given that it always leads to better
estimates, we shall continue to employ such truncation throughout the
remainder of the paper.

\oneFIG{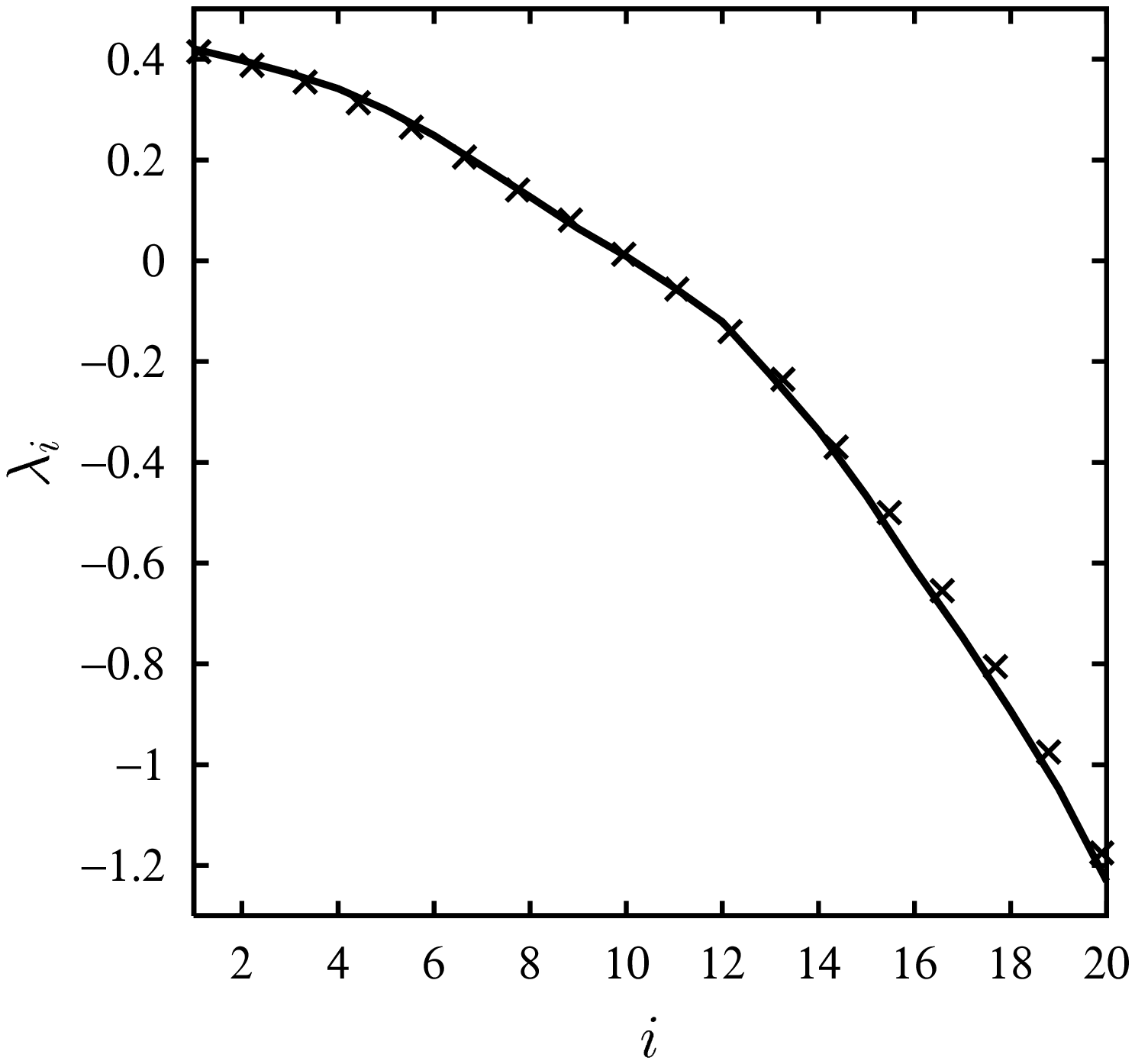}{\FigSize}{\SKEWQCAP}

Figure \ref{skewqF.ps} shows an example of the estimation of the LS
using a quasi-diagonal reconstruction with a large window  $d_s = 20$.
Compared to figures \ref{skewDT1b2F.ps} and \ref{skewDT1s20F.ps} we see
that this approach can give an excellent estimate, even though the local
map cannot be well approximated by a quadratic fit. Similar results
where obtained for different maps and coupling strengths.

\section[Estimating the LS for exponentially decaying coupling]
        {Estimating the Lyapunov spectrum for exponentially decaying
coupling \label{LS-GRALCOUPLING}}

The CML used to illustrate our method in the previous sections only allowed
interactions between nearest neighbours. There are however many other
systems of interest where the coupling acts over greater distances. In this
section we investigate the performance of our approach in such cases.
Typically, it is assumed that the coupling decreases in strength with
distance, giving a band-diagonal Jacobian with sub-diagonals whose entries
decay as we move away from the diagonal. A suitable paradigm model to
represent this is a  CML where the coupling range is in fact infinite but
whose coupling coefficients decay exponentially with distance:
\begin{equation} \label{exp-coup}
x^{n+1}_i  = {1-\beta\over 1+\beta}
\sum_{k=-\infty}^{\infty} {\beta^{|i-k|} f(x^n_{i-k})},
\end{equation}
where the coupling parameter $\beta\in(0,1)$. The limit $\beta \rightarrow 0$
corresponds to the uncoupled case, {\em i.e.}~$x^{n+1}_i$ depends only
on $x^n_{i-k}$. The limit $\beta \rightarrow 1$ corresponds to global
coupling of all the sites with the same coefficient. Thus increasing $\beta$
effectively increases the range of the coupling and results in more and
more sub-diagonals of the Jacobian becoming significant. Whilst for $\beta$
close to 0, it is reasonable to expect that  a tri-diagonal reconstruction
will still suffice, it is clear that as $\beta$ increases, more
sub-diagonals will need to be taken into account.

We thus investigated the effects of using a higher-diagonal reconstruction
for data from a lattice with a relatively large $\beta$ (0.35). If we fix
the size $q$ of the admissible interaction range we need to estimate a map
depending on $2q+1$ variables:
\begin{equation}\label{F_r-quasi}
F_r^{(q)}(x_{r-q}^n\dots,x_{r-1}^n,x_{r}^n,x_{r+1}^n,\dots,x_{r+q}^n)
=x_{r}^{n+1},
\end{equation}
%

\oneFIG{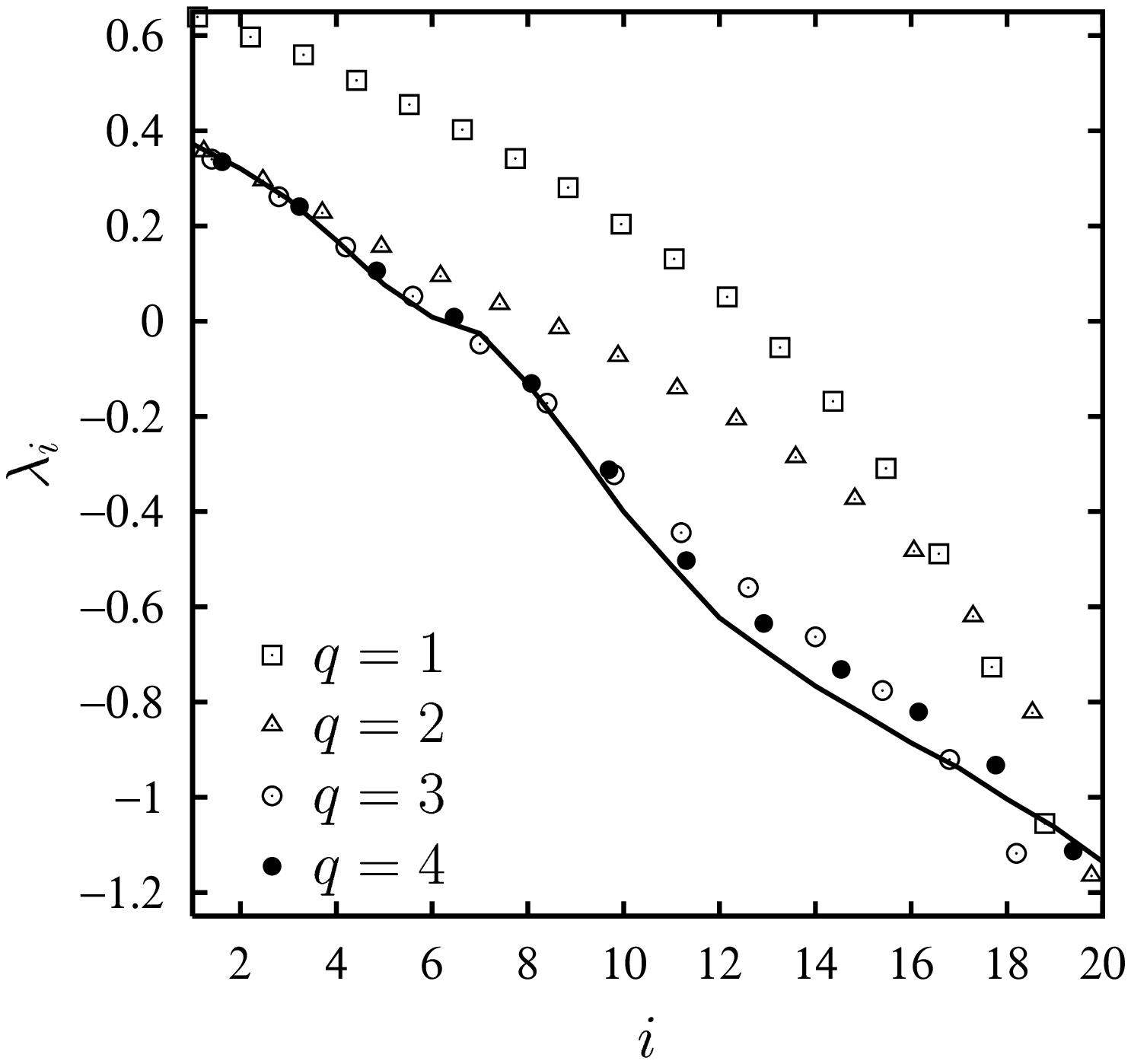}{\FigSize}{\EXPCOUPCAP}

We now need to discard $q$ outer layers of the estimated Jacobian to remove
boundary effects, and hence if our observation window is of size $d_s$, we
can estimate $d_s - 2q$ Lyapunov exponents.

Figure \ref{expcoupF.ps} shows the results of this approach applied to
the CML coupling scheme (\ref{exp-coup}) for different values of $q$
(still with a skewed logistic map for local dynamics). We see that the
tri-diagonal ($q=1$) reconstruction completely fails to approximate the
LS. However, as $q$ increases the reconstruction rapidly improves. 
For $q=2$ the largest
Lyapunov exponents are captured accurately and the estimate of the
remainder of the spectrum is still quite poor. This improves significantly 
for $q=3$ and $q=4$ where the approximation is rather good,
particularly given that the coupling is not genuinely over a finite
range.  Increasing $q$ further led to a deterioration of the estimates
of the LS. This is due to the reappearance of the ``curse of
dimensionality'': for $q\ge 5$ we have to find neighbours in dimension
$d = 2q+1 \ge 11$. We repeated these experiments for other values of
$\beta$ and other local maps, obtaining similar qualitative results.

\section{Discussion and generalisations}

We have shown that by using a quasi-diagonal reconstruction of the
Jacobian we are able to estimate the LS of a variety of CML's using only
time series data with a reasonable degree of accuracy. The most
difficult test was where the local dynamics was given by a skewed
logistic map and the coupling was exponentially decaying. In that case, we 
had to take a 9-diagonal reconstruction of the Jacobian ({\em i.e.}~$q=4$). 
Increasing $q$ beyond this leads to increasing errors, due to
the increase in the dimensionality of the reconstruction space that we
use. We therefore have a dichotomy when trying to estimate LS for
systems with extended interactions: on the one hand we would like to
include as many sub-diagonals as possible ({\em i.e.}~$q$ as large as
possible), however, on the other hand, $q$ cannot be chosen too large
($q\leq 5$) otherwise we again encounter the ``curse of dimensionality''.

Of course, in practice if we do not know the dynamics of a system, we are
unlikely to know {\em a priori} the range of the coupling. However, this is
something that can be estimated from a multivariate time series, using for
example some kind of cross correlation. This in turn will allow us to
estimate the width of the band of the Jacobian that is most appropriate to
capture the essence of these interactions.

A related problem arises when trying to reconstruct LS from multivariate
time series produced by systems with a continuous space variable, or
when probing real life extended dynamical systems. Consider for example
a one-dimensional PDE and assume that we are free to choose the sampling
intervals in both time and space at which data is observed. The Jacobian
will contain a different number of significant sub-diagonals depending
on the choice of these intervals. This is illustrated in figure
\ref{meshpde.eps} where we show the so called cone-horizon for the
evolution of a perturbation in space-time. The cone-horizon corresponds
to the region of space-time where a perturbation, applied at the summit
of the cone, can influence downstream positions. Any point outside the
cone-horizon cannot ``feel'' the presence of the disturbance. Such
cone-horizons always exist for extended dynamical systems where the
interactions have finite range, {\em i.e.}~where information propagation
has finite speed. Now suppose that we observe the system at regular
intervals in space and time. This corresponds to the the intersections
of the grid shown in the figure. Let us denote by $\tau$ the temporal
sampling; for simplicity we assume the spatial sampling interval is
fixed, since varying $\tau$ is sufficient to demonstrate our point. For
the example depicted in the figure a sampling interval of $\tau$ yields
a tri-diagonal Jacobian. This is  because there are only three sites in
the cone-horizon after a time $\tau$. Thus, for this choice of
space-time discretisation it should be sufficient to use a tri-diagonal
reconstruction of the Jacobian, {\em i.e.}~with $q=1$. However, if we
double the sampling interval to $2\tau$ we shall need to use a
5-diagonal reconstruction in order to include the 5 sites in the
cone-horizon after a time $2\tau$.

\oneFIG{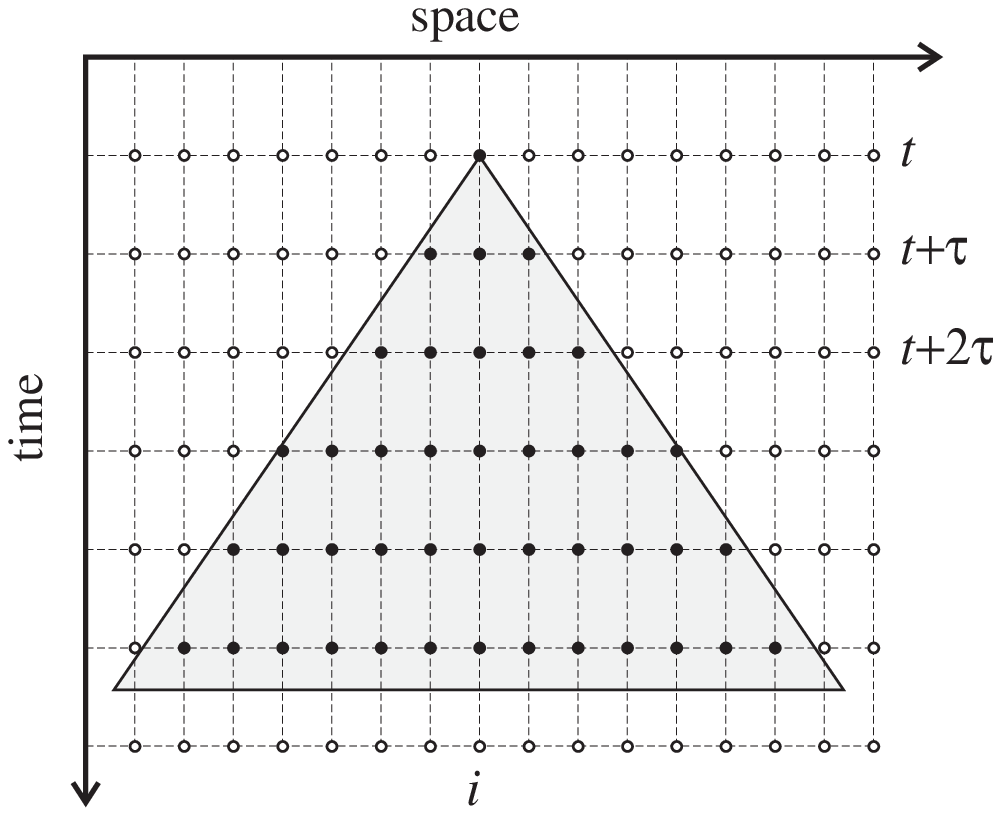}{\FigSize}{\MESHPDECAP}

We intend to investigate the application of the quasi-diagonal method
to the estimation of a LS from a time series generated by a PDE in a future
paper. Additionally, we have hitherto assumed that the observable is the
actual state variable at a site. This is unlikely to be the case in many
practical applications. In such a case, we expect that the best approach
would be a combination of the quasi-diagonal reconstruction method
presented here with a local spatio-temporal reconstruction of the dynamics.

We would like to thank D.S.~Broomhead, J.~Huke and T.~Schreiber
for useful discussions. This work was carried out under a UK Engineering
and Physical Sciences Research Council grant (GR/L42513). JS would also
like to thank the Royal Society, the Leverhume Trust and the Royal
Commission for the Exhibition of 1851 for financial support.


\end{document}